\documentclass[aps,prl,twocolumn,longbibliography]{revtex4-2}

\usepackage{mathrsfs}
\usepackage{tabularx}
\usepackage{slashed}
\usepackage{amsmath}
\usepackage{amsfonts}
\usepackage{slashed}
\usepackage{bm}
\usepackage{graphicx,color}
\usepackage[dvipsnames]{xcolor}
\usepackage{times}
\usepackage{braket}
\usepackage{orcidlink}
\usepackage[caption=false]{subfig}
\AtBeginDocument{%
   ~\newwrite\bibnotes
   ~\def\bibnotesext{Notes.bib}
   ~\immediate\openout\bibnotes=\jobname\bibnotesext
   ~\immediate\write\bibnotes{@CONTROL{REVTEX41Control}}
   ~\immediate\write\bibnotes{@CONTROL{%
    apsrev42Control,author="08",editor="1",pages="1",title="0",year="1"}}
    ~\if@filesw
    ~\immediate\write\@auxout{\string\citation{apsrev42Control}}%
   ~\fi
}%

\begin{document}

\title{Symmetry-Fractionalized Skin Effects in Non-Hermitian Luttinger Liquids}

\author{Christopher Ekman\orcidlink{0009-0003-6426-2376}}\thanks{christopher.ekman@fysik.su.se}

\author{Emil J. Bergholtz\orcidlink{0000-0002-9739-2930}}\thanks{emil.bergholtz@fysik.su.se}
\affiliation{Department of Physics, Stockholm University, SE-106 91 Stockholm, Sweden}

\author{Paolo Molignini\orcidlink{0000-0001-6294-3416}}\thanks{paolo.s.molignini@jyu.fi}
\affiliation{Department of Physics and Nanoscience Center, University of Jyväskylä, P.O. Box 35 (YFL), University of Jyväskylä, FI-40014 Jyväskylä, Finland}
\affiliation{Department of Physics, Stockholm University, SE-106 91 Stockholm, Sweden}

\date{\today}

\begin{abstract}
In one dimension, strongly correlated gapless systems are highly constrained due to conformal invariance, leading to the decoupling of low energy degrees of freedom corresponding to different symmetry sectors. 
The most familiar example of this is spin-charge separation. 
Here, we extend this mechanism to the non-Hermitian realm by demonstrating that skin effects corresponding to different symmetry sectors exhibit an emergent decoupling. We establish this for $N$ flavor fermions and demonstrate it numerically for the special case of the Hubbard model, in which spin and charge skin effects separate at low energies. 
Finally, we construct an interaction-enabled $E_8$  skin effect with no free fermion counterpart.
\end{abstract}

\maketitle
{\it Introduction.--}
Relaxing Hermiticity in quantum systems leads to new localization phenomena that defy conventional rules~\cite{Ashida:2020,Bergholtz:2021,Okuma:2023}.
The non-Hermitian skin effect (NHSE) is arguably one of the most famous localization effects displayed by effective non-Hermitian Hamiltonians \cite{Yao:2018,Kunst:2018,Lee2016PRL,Martine2018PRB,Okuma:2020,Gong:2018} and Liovillians \cite{Song:2019,Haga2021,FanYang2022}.
In the NHSE, an extensive number of eigenstates accumulate at the boundaries of a finite system under open boundary conditions.
First identified in a one-dimensional (1D) model with non-reciprocal hopping by Hatano and Nelson~\cite{Hatano:1996}, the NHSE has since been understood as a breakdown of conventional bulk–boundary correspondence~\cite{Yao:2018,Kunst:2018,Yokomizo:2019,Okuma:2020}, e.g. described using biorthogonal or non-Bloch band theory and discovered in many other settings~\cite{Zhou:2018,Longhi:2019,Lee:2019,Lee:2019-2,Song:2019,Zhang:2020,Edvardsson2020,Li:2020,Li:2020-2,Yi:2020,Borgnia:2020,Zirnstein2021,Claes:2021,Longhi:2022,Wang:2023,Molignini:2023,Denner:2024,Yang:2025}.
Comprehensive classifications of non-Hermitian topological phases have been formulated~\cite{Gong:2018,Kawabata:2019,Yang_nHHomotopy_RepProgPhys2024}, but in this framework the NHSE is primarily viewed as a single-particle spectral instability, and the role of interactions remains less clear. 

Interactions can, however, coexist with non-Hermitian descriptions in driven-dissipative experiments, such as in cold atomic gases with controlled loss~\cite{Gou_2020,Liang_2022,Li:2019}, photonic lattices exhibiting the optical Kerr effect along with engineered gain and absorption~\cite{Imamoglu_1997,Mukherjee_2020,Ruter:2010,Zhao:2019,Weidemann:2020}, and topoelectrical circuits~\cite{Fitzpatrick:2017, Helbig:2020}.
This highlights a pressing need for a theoretical description that incorporates many-body effects.
In fact, several recent works have explored classification extensions to interacting settings~\cite{Kawabata:2022} and reported many-body skin effects~\cite{Hamazaki:2019,Alsallom:2022,Zhang:2022,Kim:2024,Yoshida:2024,Qin:2026,ekman_2024,wang_2025,PhysRevB.102.235151,roccati2026perspective,Mao_2024}.
However, a universal low-energy framework for NHSE in interacting systems is still lacking.
In particular, it remains unclear how boundary localization reorganizes in the presence of interacting symmetry sectors, or how non-reciprocity competes with or enhances interaction-driven phenomena.

These questions are particularly enticing in one dimension (1D). 
In Hermitian systems, the low-energy behavior of generic locally-interacting 1D fermions is universally described by Luttinger liquid (LL) theory, as first elucidated by Haldane~\cite{Haldane_1981, Giamarchibook}. 
This is typically achieved by `bosonization', where the fermionic field operators are expressed in terms of bosons. 
The remarkable predictive power of LL theory stems from the small number of parameters it requires and from its exact solvability due to conformal symmetry \cite{cftbook}. 
For fermions with multiple flavors, such as spin $1/2$ fermions, the bosonization procedure results in two decoupled Luttinger liquids -- one for the charges and one for the spins.
This means that the fermionic quantum numbers fractionalize, and spin and charge excitations may propagate at different velocities.

Given the success of LL theory, it is natural to ask whether interacting non-Hermitian systems in 1D admit a similar \textit{symmetry-resolved} low-energy description.
More precisely, we seek a framework capable of predicting how non-reciprocity impacts collective modes, fractionalized degrees of freedom, and boundary localization.

\begin{figure}[t!]
    \centering
    \includegraphics[width=0.99\columnwidth]{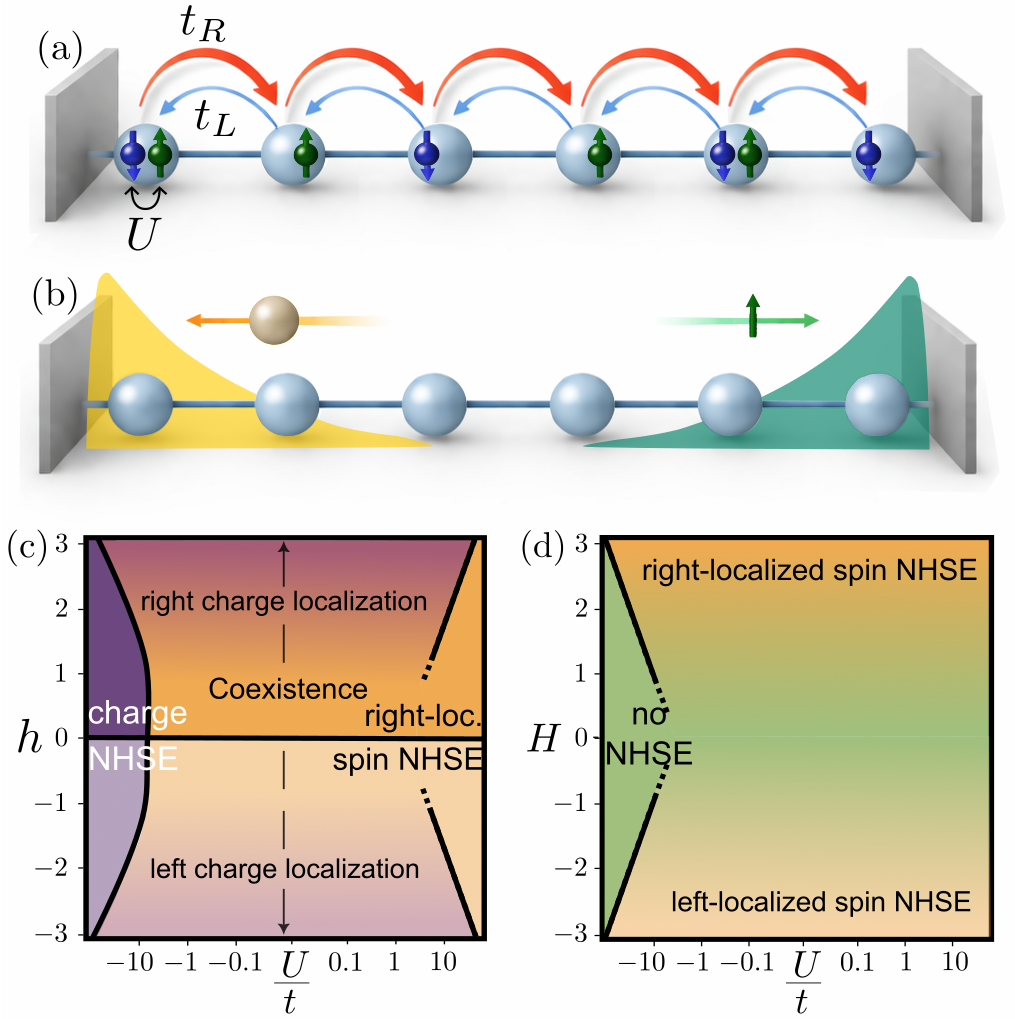}
    \caption{\textbf{The Hatano-Nelson-Hubbard model and its phase diagram.} 
    (a) Sketch of an open-ended Hatano-Nelson-Hubbard model.
    (b) Sketch of the symmetry-fractionalized skin effect, with charge and spin localizing at opposite ends.
    (c)-(d) Ground-state localization properties at half-filling extracted from the mean center of mass of the density and spin profiles, plotted as a function of gauge fields $h$, $H$ and interactions $U$. In (c), $H=1$, while in (d) $h=0$.}
    \label{fig:phase-diag}
\end{figure}

In this work, we develop a symmetry-resolved low-energy framework for interacting non-Hermitian systems in 1D.
We show that nonreciprocal hopping can be understood as coupling distinct LL sectors to constant imaginary background gauge fields, allowing boundary localization to be analyzed directly within the bosonized theory. 
For a purely linear dispersion, we demonstrate that symmetry sectors remain strictly decoupled, enabling independent and tunable skin effects for charge and spin, as illustrated in Fig.~\ref{fig:phase-diag}(b).
Corrections beyond the linear regime generically mix these sectors, enabling boundary localization to propagate between them.
We substantiate this picture in the prototypical Hatano–Nelson–Hubbard model through analytical and numerical calculations, revealing regimes of symmetry-resolved and interaction-induced skin effects. 
Finally, we construct an interaction-enabled NHSE in a system of eleven fermionic flavors whose interacting low-energy theory realizes an emergent $E_8$ sector.
This demonstrates that interactions can generate boundary localization phenomena with no free-fermion counterpart. 
Such multi-flavor systems could in principle be implemented in ultracold atoms with large hyperfine manifolds, such as alkaline-earth atoms (e.g. $^{87}$Sr) supporting many internal states \cite{Cazalilla_2014}.

{\it Symmetry-resolved skin effects.--}
Massless Dirac fermions with $N$ flavors interacting on a line can be described by the action $\mathcal{S}=\int dxdt\ \bar{\Psi} \slashed{\partial}\Psi$, where $\Psi$ is in the fundamental representation of $U(N)$.
It has $U(N)$ symmetry, and is therefore equivalent to a $U(N)_1$ WZW theory \cite{Witten_1984}. 
Close to half-filling, it can equivalently be expressed as a bosonic field theory
\begin{equation}
\mathcal{S}_b=\frac{1}{2\pi}\int dx\,dt \sum_{n=1}^N \frac{1}{K_n}\bigg[\frac{1}{v_n}(\partial_t\varphi_n)^2-v_n(\partial_x\varphi_n)^2\bigg],
\end{equation}
%
see the end matter for details on our conventions.
The charge density of each fermion flavor is $\rho_n=-\frac{1}{\pi}\partial_x\varphi_n$. 
The Luttinger parameters $K_n$ take forward-scattering interactions into account, with $K_n<1$ corresponding to attractive interactions, $K_n>1$ to repulsive interactions, and $K_n=1$ to no interactions.
The $U(N)$ symmetry remains only if all $K_n,v_n$ are the same. 

Now, let $G$ be a subalgebra of the affine Lie algebra $U(N)_1$ with $U(N)$ generators $T^a,a=1,2,..., n$ and associated bilinear currents $J^a_{\mu}=\bar{\psi}_n\gamma_{\mu}T^a_{nm}\psi_m$. 
A model exhibiting the NHSE for the conserved charges associated with $G$ may be constructed by minimally coupling $\mathcal{S}$ to a constant imaginary background gauge field $H_{\mu}=H^a_{\mu}T^a$ with gauge group $G$, leading to the fermionic action 
   $\mathcal{S}_{H}=\int dxdt\ \bar{\Psi} (\slashed{\partial}+\slashed{H})\Psi.$ 

To see that this leads to the NHSE, choose a basis of $G$ such that $H^{a}T^a$ is diagonal.
Then $\rho_{H}=\sum_{i=1}^N H_i\rho_i$, where $H_i$ are the elements along the diagonal of $H^aT^a$ and $\rho_i=\psi^{\dagger}_i\psi_i$. 
Note that $\mathrm{Tr}\ H^aT^a=\sum_iH_i=0$, unless $G=U(1)$. 
Next, take the complementary subalgebra $G'$ such that $U(N)_1=G\oplus G'$ and let $\rho_{h}=h^a J^a_0\in G'$ be a density corresponding to $G'$. 
Since $G$ and $G'$ are different subalgebras, their elements mutually commute, $[H^aT^a,h^bT^b]=0$. 
Choose the basis such that $h^bT^b$ is also a diagonal matrix with entries $h_i$. 
Then $$\langle \rho_h(x)\rangle\!=\!\sum_{i=1}^N\! h_i\! \frac{\langle e^{\!-\!\sum_{j=1}^N\!\int_{0}^Lx'H_j\rho_jdx'}\!\rho_i(x) e^{\!-\!\sum_{j=1}^N\!\int_{0}^Lx'H_j\rho_jdx'}\rangle_0}{\langle  e^{\!-2\!\sum_{j=1}^N\!\int_{0}^Lx'H_j\rho_jdx'}\rangle_0}$$ is the ground-state expectation value of $\rho_h$. Here the subscript $0$ indicates that the expectation values are taken with respect to the theory where the imaginary gauge fields are $0$. It can be obtained via a similarity transform, as explained in the supplemental material \cite{supmat}.
For a single flavor, this was computed in Ref.~\cite{Dora_2023}. 
Here, the expectation value evaluates to~\cite{supmat}
\begin{equation}\label{h-dens}
    \langle \rho_h(x)\rangle=2\sum_{i=1}^N \frac{h_iH_i K_i}{\pi^2}\log\left|\tan\frac{\pi x}{2L}\right|.
\end{equation}

The edge at which $\rho_h$ localizes is determined by the sign of $\sum_{i=1}^Nh_iH_iK_i$.
In particular, it is positive for $\rho_H$ but may take either sign for $\rho_h$. 

Now suppose that all $K_i=K$ so that the $U(N)$ symmetry is preserved. 
Then if $\rho_H$ and $\rho_h$ are charges corresponding to different symmetry sectors, the basis may be chosen so that $h_i=0$ whenever $H_i\neq 0$ and vice versa, so that $\sum_{i}h_iH_iK=0$, and hence equation \eqref{h-dens} gives $\langle\rho_h(x)\rangle=0$. 
This shows that for a model with a purely linear dispersion, ground state skin effects corresponding to different symmetry sectors \textit{always} decouple. 

This result crucially depends on the linear dispersion of the Dirac fermions. 
If the dispersion is not linear, curvature terms, such as $-\frac{1}{2m}\psi^{\dagger}\partial_{xx}\psi$, modify the bosonized theory by the addition of irrelevant operators $\sim(\partial_x\varphi_{\pm})^3$ which mix the different symmetry sectors  \cite{Haldane_1981,Imambekov2012}. 
Therefore, if there is a constant imaginary background gauge field $H_{\mu}$ corresponding to one of the symmetry sectors, say $G$, there may be skin effects in the other symmetry sector $G'$, as will be discussed in example 1 below. 

The skin effects in $G'$ can be suppressed by gapping out the modes corresponding to it. 
Therefore, a model exhibiting the skin effect only within the $G$ sector may be constructed by taking $U(N)$ fermions, coupling them to an imaginary gauge field, and gapping out the $G'$ sector. 
In the following sections we demonstrate this procedure in two examples: the Hubbard model, which has $U(2)=SU(2)\otimes U(1)$ symmetry, and a field-theoretical model exhibiting an intrinsically strongly coupled skin effect corresponding to an $E_8$ symmetry.

{\it Example 1: The Hatano-Nelson-Hubbard model.--} 
The simplest realization of the above considerations is provided by the Hatano-Nelson-Hubbard model sketched in Fig.~\ref{fig:phase-diag}(a):
%
%
%
\begin{align} \label{HNH}
    \mathcal{H} =-&\sum_{j=1}^L\left(t_R \Psi_{j+1}^{\dagger}\Psi_j+t_L\Psi_{j}^{\dagger}\Psi_{j+1} + U n_{j,\uparrow}n_{j,\downarrow}\right)
\end{align}
where $t_R=t e^{h+H\sigma^z}$, $t_L=t e^{-h-H\sigma^z}$ are the hopping amplitudes, $L$ is the number of sites, $\Psi_j=(c_{j,\uparrow}\ c_{j,\downarrow})^T$ is a vector containing the fermionic annihilation operators for up and down spins at each site and $n_{j,s} \equiv c_{j,s}^{\dagger} c_{j,s}$, $U$ is the interaction strength, and $h,H$ parametrize the non-reciprocity. 
The Hubbard model is recovered in the Hermitian limit $h=H=0$. Conversely, the non-Hermitian model may be obtained from the Hubbard model by introducing a pair of imaginary gauge fields. The $U(1)$ symmetry is gauged by adding a field $A_{\mu}$ which couples to the charge, and the $SU(2)$ symmetry by a field $B_{\mu}=\sum_{i=1}^3B_{\mu}^i\sigma^i$. Letting the $\mu=0$ component of both of these fields be $0$ and taking $A_1=ih,B_1=iH\sigma^z$ the Hatano-Nelson-Hubbard model is obtained.

The continuum limit of this model is given by a pair of scalar bosons, one corresponding to the $U(1)$ charge sector, and the other corresponding to the $SU(2)$ spin sector. 
Using abelian bosonization for the spin up and spin down fields, and defining $\phi_c=\frac{1}{\sqrt{2}}(\varphi_{\uparrow}+\varphi_{\downarrow}),\ \phi_{\sigma}=\frac{1}{\sqrt{2}}(\varphi_{\uparrow}-\varphi_{\downarrow})$, the Hamiltonian density decouples into charge and spin parts as $\mathcal{H}=\mathcal{H}_c+\mathcal{H}_{\sigma}$ where
\begin{align}
\mathcal{H}_j&=\frac{v_j}{2\pi}\left(\frac{1}{K_j}(\Pi_j+i\beta_j)^2+K_j(\partial_x\phi_j)^2\right)+U_j\cos(2\sqrt{2}\phi_j)\nonumber
\end{align}
where $j=c,\sigma$, $\beta_c=h,\beta_{\sigma}=H$, and the term proportional to $U_c$ is present only at half-filling due to Umklapp processes. The general form of the low-energy theory is valid throughout the regime where the low-energy bosonized description applies, with the Luttinger parameters given in \cite{supmat}, (see also Ref. ~\cite{Moreno_2011,Ogata_1991}).

The two sine-gordon couplings proportional to $U_c,U_{\sigma}$ can gap out the charge- and spin degrees of freedom, respectively, if they are relevant. 
For $U_c$ this corresponds to any $U>0$ while for $U_{\sigma}$ degree of freedom it corresponds to any $U<0$. 
At half-filling, the charge degree of freedom is therefore gapped for any $U>0$ while the spin degree of freedom is gapped for any $U<0$. 
In the former case we expect there to only be a spin skin effect, while in the latter case we expect there to only be a charge skin effect. This can be seen explicitly in the strong coupling limits, which are treated in the end matter, and it will be further demonstrated numerically below. 
Note that in this model, taking $U\rightarrow -U$ amounts to exchanging the spin and charge degrees of freedom \cite{Essler_2005}. 

Away from half-filling particle-hole symmetry does not hold. In the bosonized picture, a particle-hole transformation acts as $\phi_c\rightarrow-\phi_c$, so away from half-filling irrelevant spin-charge mixing terms such as $\lambda\partial_x\phi_c(v_{\sigma}^2(\partial_x\phi_{\sigma})^2-(\partial_t\phi_{\sigma})^2)$ are permitted by symmetry, and are therefore generically present in the low-energy theory. Such terms may cause the skin-effects of different symmetry sectors to couple, potentially resulting in edge localization in symmetry sectors where there is no imaginary gauge field. An example of this is provided in the Supplemental Material \cite{supmat}. In this case, an imaginary $SU(2)$ gauge field leads to a non-trivial $\sigma$ background, which effectively leads to a position-dependent chemical potential for the charge degree of freedom.

{\it Numerical results --}
We now demonstrate the accuracy of our symmetry-fractionalized picture using exact diagonalization (ED) calculations of the Hatano--Nelson--Hubbard chain~\eqref{HNH} under open boundary conditions.
We obtain the ground state properties  following an Arnoldi iteration scheme performed in the EDITH software~\cite{EDITH}. 
In particular, we compute the ground-state density and spin profiles -- $\langle n_j\rangle$ and $\langle S^z_j\rangle$ -- to characterize boundary accumulation through their normalized mean centers of mass (mcom).
The mcom takes the value $\simeq 1/2$ for extended profiles and approaches $0$ or $1$ for left- or right-localized skin modes, respectively.

Based on the ground-state observables, we qualitatively summarize the localization properties as functions of interaction strength $U$ and the two imaginary gauge fields $h$ and $H$ in the phase diagram of Fig.~\ref{fig:phase-diag}(c)-(d).
The raw heatmaps obtained from the mcom are reported in the Supplemental Material~\cite{supmat}, where we scan $U/t$ between $-\exp(4)\approx 54.6$ and $+54.6$ and $h,H$ between $-3$ and $+3$
The degrees of freedom split into a spin sector corresponding to $SU(2)$ and a charge sector corresponding to $U(1)$, which can have independent skin effects and may develop gaps independently. For strong repulsive interactions, $U/t\gg0$, the charge sector has a gap at half-filling, which suppresses its skin effect [orange region in Fig.~\ref{fig:phase-diag}(c)]. Conversely, for strong attractive interactions $U/t\ll 0$, the spin sector has a gap, which suppresses its skin effect [purple region in Fig.~\ref{fig:phase-diag}(c)].

Note that in contrast to the field theoretical model in the preceding section, the phase boundaries obtained numerically depend on the imaginary gauge fields. To see why, we note that the field theory describes the renormalization group fixed points, in which all microscopic details have been hidden. Such details can be taken into account by including irrelevant operators, which lead to changes in the precise form of the phase transition lines, without adding or removing any phases or phase transitions.

\begin{figure}[t!]
    \centering
    \includegraphics[width=\columnwidth]{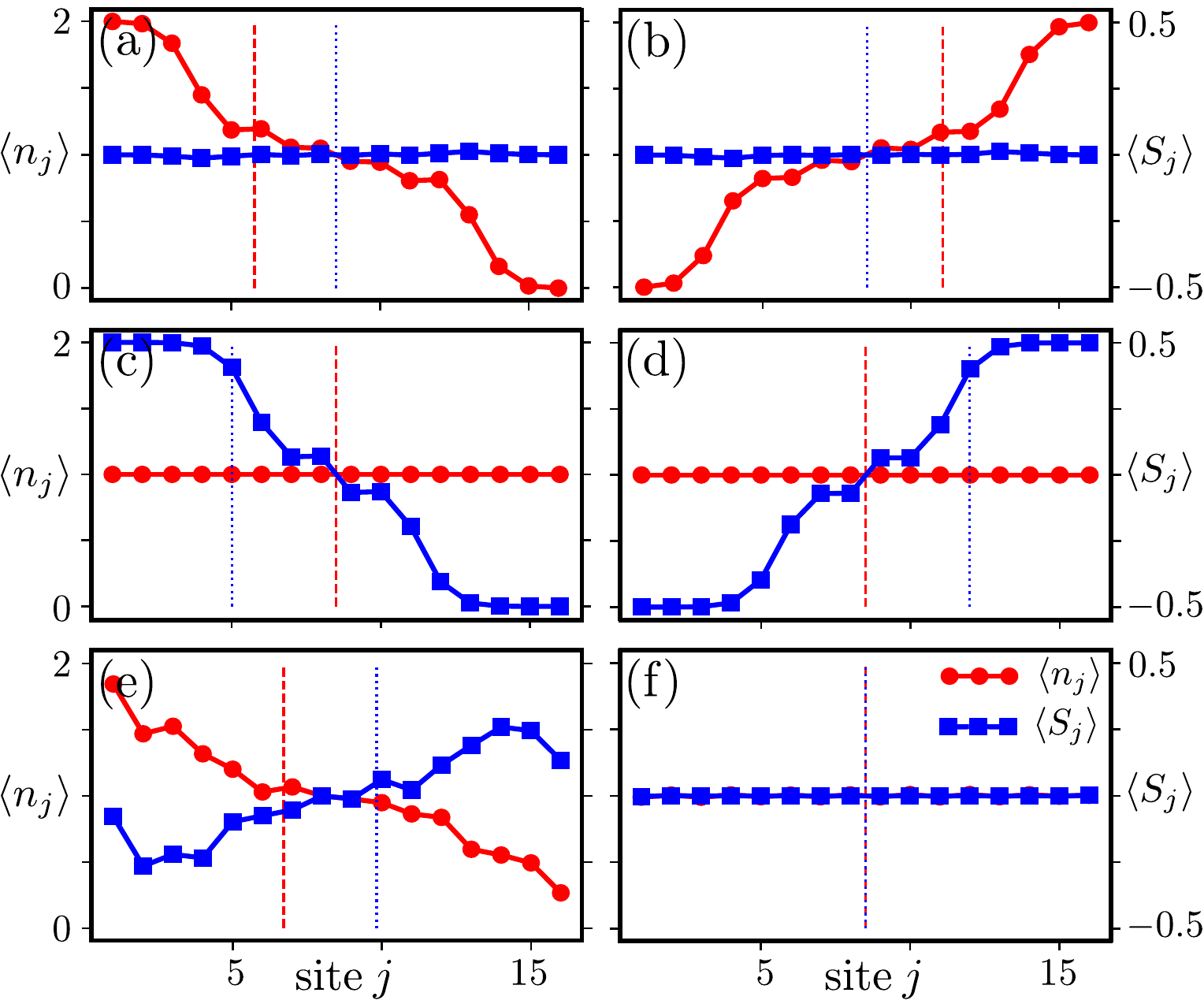}
    \caption{\textbf{Charge and spin skin effect separation.} Ground-state density and spin expectation value for a system of $L=16$ sites for different choices of gauge fields $h$, $H$ and interaction strength $U$.
    (a) $h=-2$, $H=1$, $U/t=-55$.
    (b) $h=2$, $H=1$, $U/t=-55$.
    (c) $h=0$, $H=-2$, $U/t=55$.
    (d) $h=0$, $H=2$, $U/t=55$.
    (e) $h=-2$, $H=1$, $U/t=-0.02$.
    (f) $h=0$, $H=1$, $U/t=-55$.
    The vertical lines indicate the mean center of mass for each curve.
    }
    \label{fig:profiles}
\end{figure}

Changing the signs of $h$ and $H$ selectively controls the boundary at which the charge and spin skin effect occur, respectively. 
The former is indicated in Fig.~\ref{fig:phase-diag}(c) and the latter in Fig.~\ref{fig:phase-diag}(d). 
In addition to the regimes where only one of the skin effects is present, there is also an intermediate LL region where both are present, and can be tuned independently.
This is further shown quantitatively in Fig.~\ref{fig:profiles}, with representative charge and spin profiles (vertical lines mark the mcom). 
Panels ~\ref{fig:profiles}(a) and ~\ref{fig:profiles}(b) show the strongly attractive region with $U/t=e^4\approx 55$, and fix $H=1$ while the sign of $h$ is reversed, causing the charge skin effect to localize on opposite sides while the skin profile remains flat. 
Panels ~\ref{fig:profiles}(c) and ~\ref{fig:profiles}(d) feature the strongly repulsive region with $h=0$ and $H=\mp 2$.
Here the charge density remains flat while the spin localization reverses. 
In panel ~\ref{fig:profiles}(e), the weakly interacting regime is shown, where both skin effects are present and mix to some extent due to the band curvature. 
Finally, panel ~\ref{fig:profiles}(f) highlights a regime where neither sector localizes, despite $H\neq 0$.

\begin{figure}[t!]
    \centering
    \includegraphics[width=\columnwidth]{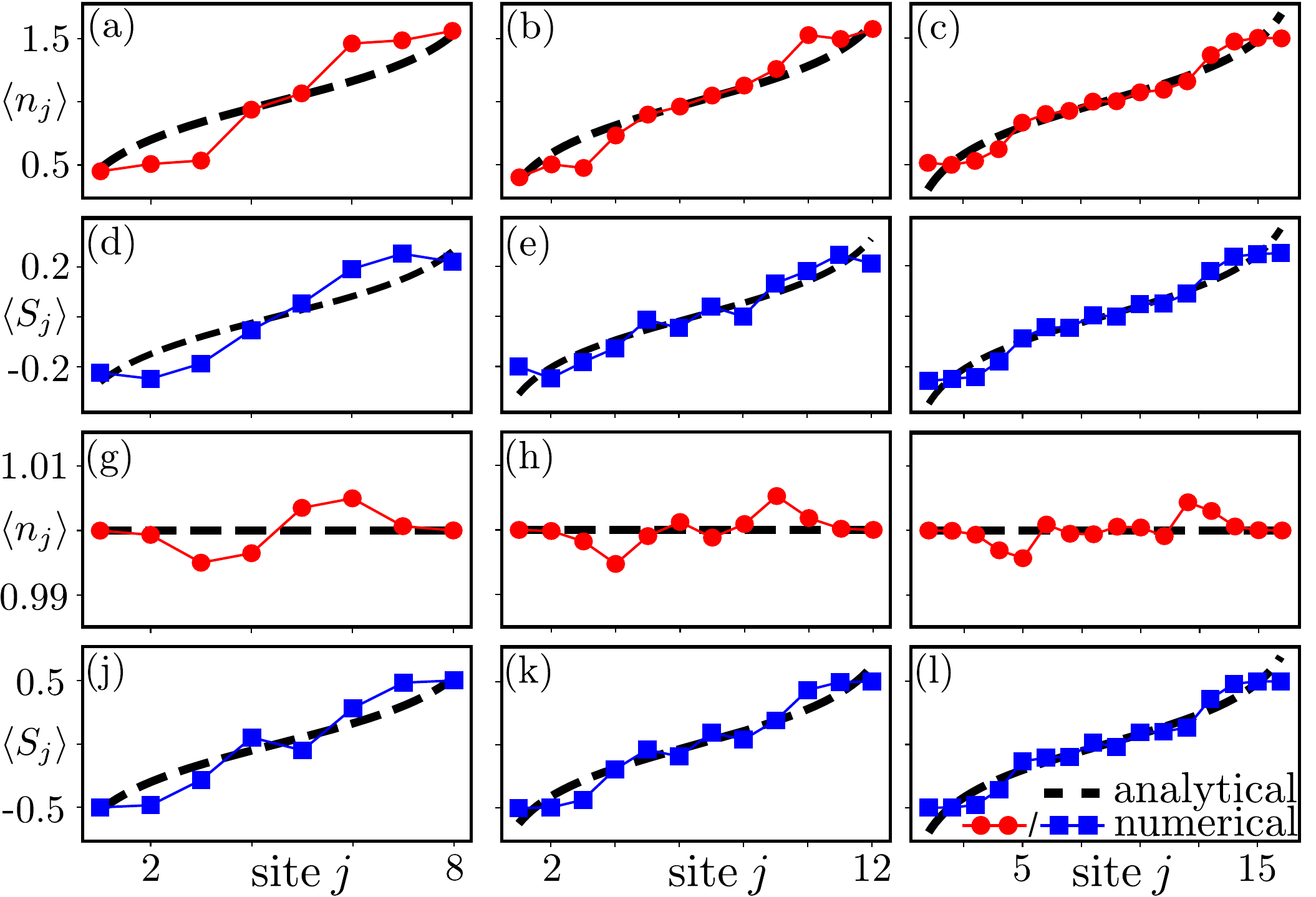}
    \caption{\textbf{Comparison between analytical prediction and numerical calculations.} 
    Each panel shows the density or spin profile obtained analytically (dashed black line) and numerically (connected circles/squares) for $h=H=1$ and increasing system size --- $L=8$ (left), $L=12$ (center), $L=16$ (right).
    (a)-(c): density profile for $U/t=0.01$.
    (d)-(f): spin profile for $U/t=0.01$.
    (g)-(i): density profile for $U/t=55$.
    (j)-(l): spin profile for $U/t=55$.
    }
    \label{fig:comparison}
\end{figure}

To draw a direct and quantitative parallel with our theoretical predictions, Fig.~\ref{fig:comparison} compares the ED results to the analytical formula of the non-Hermitian LL framework for $H=h=1$ at increasing system sizes.
The analytical profiles (dashed black lines) capture both the spatial dependence and the direction of localization observed in the numerical data (red dots/blue squares).
For weak interactions, $U/t=0.01$ [Figs.~\ref{fig:comparison}(a)--(f)], charge and spin profiles follow the predicted smooth boundary accumulation with very good agreement, improving with system size.
Remarkably, strong coupling does not invalidate the symmetry-resolved description either: for $U/t=55$ [Figs.~\ref{fig:comparison}(g)--(l)], the numerics reproduce the analytically predicted localization pattern and its relative magnitude in both sectors -- again with increasing accuracy as the system is scaled to larger sizes.
We remark that, due to the 1:1 mapping mentioned earlier between the strongly repulsive and the strongly attractive regime, the results for $U/t=-55$ are essentially equivalent to the ones already presented in Figs.~\ref{fig:comparison}(g)--(l). 
The consistency across all interactions and system sizes further supports the interpretation that, over a broad parameter range, the interacting NHSE is governed by symmetry-fractionalized low-energy physics, with interactions primarily renormalizing effective parameters while preserving symmetry structure.

{\it Example 2: The $E_8$ skin effect.--}
So far we have seen that the general recipe outlined above can be utilized to construct theories with skin effects related to unitary and special unitary symmetries.
Here we provide a more non-trivial example of a skin effect associated with $E_8$ symmetry. 
The $E_8$ Lie group is one of the exceptional simple Lie groups \cite{supmat}. 
It has rank $8$, implying that there are $8$ commuting charges, each with its own skin effect. 
Since $N$ free Dirac fermions have $U(N)$ symmetry, a theory with only $E_8$ symmetry cannot be constructed without interactions.
In order to have an $E_8$ skin effect and nothing else, all other symmetry sectors must be suppressed, since they may couple to the $E_8$ sector by irrelevant operators.
Following \cite{Lopes_2019,Lim_2023}, we note that $U(11)_1=(E_8)_1\otimes U(3)_1$, so that an $E_8$ sector may be singled out by gapping out the $U(3)$ sector.
Let $\Phi$ be the $8$-element vector of bosonic fields associated $(E_8)_1$, and $\phi$ be the $3$-element vector of fields associated with $U(3)_1$ so that $\partial_x\Phi_i$ form the Cartan subalgebra of $(E_8)_1$. 
After performing a unitary transformation \cite{supmat}, the two symmetry sectors decouple as
\begin{align}
    \mathcal{S}=-\frac{1}{2\pi}\int dxdt\ \left(\Phi^T K_{E_8}^{-1}\Box\Phi-\frac{1}{ K}\phi^T\Box\phi\right)
\end{align}
where $\Box$ is the d'Alembertian and $K_{E_8}$ is the Cartan matrix of $E_8$ which appears in the action due to the requirement that the currents are local after the transformation. 
Forward-scattering interactions have been included to place the theory at a fixed point where the two sectors are fully decoupled.
The $U(3)$ sector may be gapped out by backscattering interactions $g\sum_{i=1}^3\cos(2\sqrt{2pi} \phi_i)$, which are only there at particular fillings \cite{supmat}. 
These interactions become relevant for $K>1$, and cause a mass gap to develop. 
This happens for a range of parameters, even if additional forward scattering interactions that break the exact decoupling are included. 
The theory therefore describes a phase in which the low-energy sector has only $E_8$ symmetry. 
Next, the NHSE may be included by minimally coupling $\Phi$ to an imaginary $E_8$ vector potential $B$, which leads to a contribution $\sum_{i=1}^8B_i\partial_t\Phi_i$ to the action. 
This localizes the $E_8$ charges at the edge according to \eqref{h-dens}, depending on the choice of $B_i$.

{\it Conclusions.--}
We have developed a framework to study the interplay between non-Hermitian localization phenomena and strong interactions. 
This was done by utilizing the connection between the NHSE and gauge symmetries to write down effective low-energy theories within the framework of bosonization. 
For 1D systems, this permitted us to study how interactions can fractionalize the NHSE into different symmetry sectors. 
This was exemplified both analytically and numerically in the Hatano-Nelson-Hubbard model. 
Furthermore, we predicted a necessarily strongly correlated type of skin effect in systems of electrons with eleven flavors.

There are many possible extensions and applications of our work. Most pertinently, our framework is based on the use of non-Hermitian Hamiltonians. 
In genuinely quantum-mechanical systems, such as open quantum systems, a more accurate description is offered by the Lindblad equation. 
It would be particularly interesting to check how the strongly correlated skin effects manifest in Lindbladian time evolution, either in the dynamics or as steady-state skin effects.

The correspondence between skin effects and imaginary gauge fields beyond the $U(1)$ field is not limited to 1D, and should work in higher dimensions as well. 
There, however, different symmetry sectors do not decouple, but they can nonetheless be gapped out, leading to higher dimensional strongly correlated skin effects.

Our predictions could also be tested in experiments, with ultracold atoms in optical lattices being the most natural platform. 
As mentioned in the introduction, atoms such as $^{87}$Sr have a high degree of internal symmetry, and could realize the highly symmetrical Luttinger liquids needed to see symmetry-fractionalized skin effects. 
The technical difficulty would be to engineer dissipation such that only some of the internal states experience it.

Taken together, our results show that interactions can fractionalize the non-Hermitian skin effect into symmetry sectors, selectively suppress or isolate it by gapping complementary modes, and even enable skin effects with no free-fermion counterpart.

\begin{acknowledgments}
\paragraph{Acknowledgments.} 
C.E. and E.J.B. are supported by the Knut and Alice Wallenberg Foundation (2023.0256), and the Göran Gustafsson Foundation for Research in Natural Sciences and Medicine.
P.M. acknowledges  support by the Swedish Research Council (2024-05213).
Computation time on the Sunrise Compute Cluster of Stockholm University, on the Dardel cluster at PDC-NAISS under compute grant NAISS 2025/5-578, and on the Euler cluster at the High-Performance Computing Center of ETH Zurich is gratefully acknowledged. The exact diagonalization data is accessible on \href{https://doi.org/10.5281/zenodo.19329307}{Zenodo} and has DOI 10.5281/zenodo.19329307.
\end{acknowledgments}

\section{End matter}
\textit{Bosonization conventions.--} 
Here we provide a brief review of bosonization to fix our conventions, following \cite{Giamarchibook}. 
A chiral fermion $\psi_{r}$ with $r=\pm$ can be expressed in terms of bosons as $\psi_r=\frac{1}{\sqrt{2\pi\alpha}}e^{-i\sqrt{\pi}(r\phi(x)+\theta(x))}$, where $\alpha$ has dimensions of length, and corresponds to some microscopic length scale. 
The bosons satisfy the equal time commutation relations $[\phi(x),\partial_y\theta(y)]=i\delta(x-y)$, and the charge density and current density can then be identified as $\rho=\frac{1}{\pi}\partial_x\phi,\ j=-\frac{1}{\pi}\partial_x\theta$ respectively. 
This implies that if the fermion has a Dirac Hamiltonian $\mathcal{H}_F=iv\int dx\ \psi^{\dagger}_+\partial_x\psi_+-\psi^{\dagger}_-\partial_x\psi_-,$ where $v$ is the velocity of the fermions, then the corresponding Hamiltonian for the bosons is $\mathcal{H}_{B}=\frac{v}{2\pi}\int dxdt (\partial_x\phi)^2+(\partial_x\theta)^2$. 
Local density-density interactions have the bosonic form $\rho(x)^2\sim(\partial_x\phi)^2$, which can be taken into account by including an additional parameter $K$, yielding the Luttinger liquid Hamiltonian
$$
\mathcal{H}_{LL}=\frac{u}{2}\int dxdt\  K(\partial_x\phi)^2+\frac{1}{K}(\partial_x\theta)^2.
$$
The system described by this Hamiltonian is gapless.
More fermion flavors, as considered in the main text, can be included by adding more bosons. 
They may be coupled by forward-scattering interactions of the form $\rho_n\rho_m\sim\partial_x\phi_n\partial_x\phi_m$ where the subscripts correspond to different fermion flavors. 
Such interactions do not open a gap, and can effectively be removed by a change of basis. 
Another type of interaction that may be present due to umklapp processes or backscattering is the sine-gordon coupling $g\sin(\beta\phi)$, which opens a gap for the fermion corresponding to $\phi$, if this coupling is relevant. 
Its scaling dimension is $\Delta_g=K\beta^2/4$, so it is relevant and opens a gap if $\Delta_g<2$. 

\textit{Strong coupling limit of the Hubbard model.--}
In the strong-coupling limit, the Hubbard model reduces to the t-J model: 
\begin{align}
    H_{t-J}=-t\sum_{j=1}^N P(\Psi_{j+1}^{\dagger}\Psi_j+\mathrm{h.c.})P+\frac{4t^2}{U}\sum_{j=1}^N (\mathbf{S}_j\cdot\mathbf{S}_{j+1}) \nonumber
\end{align}
where $P$ projects onto the singly occupied subspace, while $\mathbf{S}_j=\frac{1}{2}(\sigma_j^x,\sigma_j^y,\sigma_j^z)$ are spin operators.
Minimally coupling it to a $U(1)$ gauge field $A_{\mu}$ and a $SU(2)$ gauge field $B_{\mu}$ like before we obtain the Hatano-Nelson t-J model
\begin{align}
    \mathcal{H}_{tJ}=&-t\sum_{j=1}^N P(\Psi_{j+1}^{\dagger}e^{h+H\sigma^z}\Psi_j+\Psi_{j}^{\dagger}e^{-h-H\sigma^z}\Psi_{j+1})P\nonumber\\&+\frac{4t^2}{U}\sum_{j=1}^N (\mathbf{S}_j\cdot\mathbf{S}_{j+1}+\sinh(H)\mathbf{e^z\cdot(\mathbf{S}_j\times \mathbf{S}_{j+1})})\nonumber
\end{align}
Note that the imaginary gauge fields $h$ and $H$ are the \textit{same} gauge fields as in the Hatano-Nelson-Hubbard model \eqref{HNH}. 
The continuum limit of this model is also given by a pair of scalar bosons, one corresponding to the $U(1)$ sector and the other to the $SU(2)$ sector, but the Luttinger parameters are not known analytically in general \cite{supmat}. 
At half-filling, the charge degree of freedom is projected out, resulting in a non-reciprocal Heisenberg model
\begin{equation}
    H_{Heis}=\frac{4t^2}{U}\sum_{j=1}^N(\mathbf{S}_j\cdot\mathbf{S}_{j+1}+\sinh(H)\mathbf{e^z\cdot(\mathbf{S}_j\times \mathbf{S}_{j+1})}) \nonumber
\end{equation}
which has Luttinger parameters $K=1,v_{\sigma}=\frac{2\pi v_F^2}{aU}$.

\bibliography{hn_hubbard}


\pagebreak
\newpage

\onecolumngrid
\pagebreak
\newpage
 		\renewcommand{\theequation}{S\arabic{equation}}
		\setcounter{equation}{0}
		\renewcommand{\thefigure}{S\arabic{figure}}
		\setcounter{figure}{0}
		\renewcommand{\thetable}{S\arabic{table}}
		\setcounter{table}{0}

\section{Supplemental material to 'Symmetry-Fractionalized Skin Effects in Non-Hermitian Luttinger Liquids'}


\maketitle
    The supplemental material provides derivations and details in addition to the main text. 
In section 1 we derive the analytical expression for the density profile. 
In section 2 we give the precise connection between non-reciprocal hoppings in the Hatano-Nelson-Hubbard model and the imaginary gauge fields, while in section 3 we provide the Luttinger parameters for the Hubbard model at weak coupling and in the strong coupling limits at various fillings.
In section 4 we present the numerical phase diagrams of the Hatano-Nelson-Hubbard model and explain how they are obtained.
Then, in section 5 we present numerically obtained data showing how spin-charge coupling away from half-filling can impact the skin-effect profiles.
In section 6 we discuss the shape of the phase transition line between the phase with and without skin effects in the Hatano-Nelson-Hubbard model, both from an analytical and numerical points of view. 
Finally, section 7 gives a brief review on the essentials of the $(E_8)_1$ current algebra, and in section 8 we discuss the transformation that decouples $U(11)_1$ into $(E_8)_1$ and $U(3)_1$.

\subsection{Decoupled skin effects from bosonization}
To derive the expression
\begin{equation} \label{rho_h}
    \langle \rho_h(x)\rangle=\sum_{i=1}^N h_i \frac{\langle e^{-\sum_{j=1}^N\int_{0}^L\chi H_j\rho_jd\chi}\rho_i(x) e^{-\sum_{j=1}^N\int_{0}^L\chi H_j\rho_jd\chi}\rangle_0}{\langle  e^{-2\sum_{j=1}^N\int_{0}^L\chi H_j\rho_jd\chi}\rangle_0},
\end{equation}
which was given in the main text, we use that the non-Hermitian Hamiltonian can be reduced to a Hermitian one by a similarity transform
\begin{equation}
    \mathcal{H}_{H}=V(H)\mathcal{H}_0 V^{-1}(H)
\end{equation}
where $V(H)=e^{-\sum_{j=1}^N\int_0^LxH_j\rho_jdx}$. 
The normalized ground state of the non-Hermitian Hamiltonian $\ket{\Omega}$ can therefore be expressed in terms of the ground state of the Hermitian Hamiltonian $\ket{\Omega}_0$ as $$\ket{\Omega}=\frac{V(H)\ket{\Omega}_0}{\sqrt{_0\bra{\Omega}V^{\dagger}(H)V(H)\ket{\Omega}_0}},$$ where the square root is a normalization needed because -- while $V(H)\ket{\Omega}$ is an eigenstate of $\mathcal{H}_H$ with the same eigenvalue of $\mathcal{H}_0$, it is not necessarily normalized. 
Note also that $V^{\dagger}(H)=V(H)$ by definition. 
Hence, the ground state expectation value of any operator $\mathcal{O}$ is given by
\begin{align}
    \langle\mathcal{O}\rangle&=\bra{\Omega}\mathcal{O}\ket{\Omega}\nonumber\\
    &=\frac{_0\bra{\Omega}V(H)\mathcal{O}V(H)\ket{\Omega}_0}{_0\bra{\Omega}V(H)V(H)\ket{\Omega}_0}\nonumber\\
    &=\frac{\langle e^{-\int_{0}^L\chi H_j\rho_jd\chi}\mathcal{O} e^{-\int_0^L H_i\rho_i dx}\rangle_0}{\langle e^{-2\int_{0}^L\chi H_j\rho_jd\chi} \rangle_0}.
\end{align}
Setting $\mathcal{O}=\rho_h(x)$ results in equation \eqref{rho_h}. 

To evaluate it in the case of linear dispersion, we note that $[\rho_i,\rho_j]=0$ and furthermore that for linear dispersion the Hilbert space factorizes into the Hilbert spaces corresponding to the different fermion densities $\rho_i$. 
Then the expectation value also factorizes as
\begin{align}
    \frac{\langle e^{-\sum_{j=1}^N\int_{0}^LH_j\rho_jdx}\rho_i(x) e^{-\sum_{j=1}^N\int_{0}^LH_j\rho_jdx}\rangle_0}{\langle  e^{-2\sum_{j=1}^N\int_{0}^LH_j\rho_jdx}\rangle_0}&=\frac{\langle e^{-\int_{0}^L\chi H_j\rho_jd\chi}\rho_i e^{-\int_0^L H_i\rho_i dx}\rangle_0}{\langle e^{-2\int_{0}^L\chi H_j\rho_jd\chi} \rangle_0}\prod_{j\neq i}\frac{\langle  e^{-2\int_0^L H_j\rho_j dx}\rangle_0}{\langle e^{-2\int^L_0 H_j\rho_j dx} \rangle_0}\nonumber\\
    &=\frac{\langle e^{-\int_{0}^L\chi H_j\rho_jd\chi}\rho_i e^{-\int_0^L H_i\rho_i dx}\rangle_0}{\langle e^{-2\int_{0}^L\chi H_j\rho_jd\chi} \rangle_0}.
\end{align}
This expectation value was evaluated in \cite{Dora_2023} by using the bosonization relation $\rho_i=-\frac{1}{\pi}\partial_x\phi_i$ considering derivatives of the expectation value of the vertex operator $\langle e^{i\lambda\phi_i(x)}\rangle$. 
The result is
\begin{equation}
    \frac{\langle e^{-\int_{0}^L\chi H_j\rho_jd\chi}\rho_i e^{-\int_0^L H_i\rho_i dx}\rangle_0}{\langle e^{-2\int_{0}^L\chi H_j\rho_jd\chi} \rangle_0}=\frac{2H_i K_i}{\pi^2} \log|\tan(\frac{\pi x}{2L})|
\end{equation}
where $K_i$ is one of the Luttinger parameters corresponding to $\rho_i$.
Inserting this into the expression for $\langle \rho_h(x)\rangle$ we obtain
\begin{equation}
    \langle \rho_h(x)\rangle=2\sum_{i=1}^N \frac{h_iH_i K_i}{\pi^2}\log\left|\tan\frac{\pi x}{2L}\right|,
\end{equation}
which is the relation used in the main text.

\subsection{Additional details on the Hatano-Nelson-Hubbard model}
The Hatano-Nelson Hubbard model may be expressed explicitly as
\begin{align}
    H=-\sum_{j=1}^N t_{R\uparrow}c^{\dagger}_{j+1,\uparrow}c_{j,\uparrow}+t_{L\uparrow}c^{\dagger}_{j,\uparrow}c_{j+1,\uparrow}+t_{R\downarrow}c^{\dagger}_{j+1,\downarrow}c_{j,\downarrow}+t_{L\downarrow}c^{\dagger}_{j,\downarrow}c_{j+1,\downarrow}+U\sum_{j=1}^N n_{j\uparrow}n_{j\downarrow}
\end{align}
where the hopping amplitudes are expressed in terms of the imaginary $U(1)$ and $SU(2)$ gauge fields discussed in the main text as
\begin{align*}
    t_{R\uparrow}&=te^{h+H}\\
    t_{L\uparrow}&=te^{-h-H}\\
    t_{R\downarrow}&=te^{h-H}\\
    t_{L\downarrow}&=te^{-h+H}.
\end{align*}
The gauge fields may be obtained (nonuniquely) from the non-reciprocal hoppings as
\begin{align*}
    t&=\sqrt{t_{R\uparrow}t_{L\uparrow}}\\
    h&=\frac{1}{2}\ln\left(\frac{t_{R\uparrow}t_{R\downarrow}}{t^2}\right)\\
    H&=\frac{1}{2}\ln\left(\frac{t_{R\uparrow}t_{L\downarrow}}{t^2}\right)
\end{align*}
%

\subsection{Luttinger parameters of the 1d Hubbard model in various limits}
The Luttinger parameters of the Hermitian Hubbard model may be conveniently summarized in a table, as is done below. The analytic expressions and their derivations can be found in e.g. \cite{Essler_2005}. 
In it, the Luttinger parameters are stated in terms of the Fermi velocity $v_F$ and the on-site interaction $U$. 
The charge and spin $K$-parameters are $K_c$ and $K_{\sigma}$ respectively, while the charge and spin excitation velocities are $u_c$ and $u_{\sigma}$ respectively. 
It is always assumed that there is an equal number of up and down spins. 
The attractive case can be mapped to the repulsive case, which exchanges the spin and charge degrees of freedom \cite{Essler_2005}, as is manifest in the table. 

\begin{center}
    \begin{tabular}{|c|c|c|c|c|}
    \hline
     region & $K_c$ & $K_{\sigma}$ &$u_c$& $u_{\sigma}$\\
     \hline
     $\frac{U}{\pi v_F}\approx0$&  $\sqrt{1+\frac{U}{\pi v_F}}$ &$\sqrt{1-\frac{U}{\pi v_F}}$& $v_c=v_FK_c$  &$v_{\sigma}=v_FK_{\sigma}$ \\
     \hline
     $\frac{U}{\pi v_F}\gg1, k_F\neq\frac{\pi}{2}$ & $\gtrsim 0.5$\cite{Moreno_2011, Ogata_1991} & $1$ & $\sim v_F$ & $\sim \frac{2\pi v_F^2}{aU}$\\
     \hline
     $\frac{U}{\pi v_F} \gg 1, k_F=\frac{\pi}{2}$ & none & $1$ & none & $\frac{2\pi v_F^2}{aU}$\\ \hline
     $\frac{U}{\pi v_F}\ll-1, k_F\neq\frac{\pi}{2}$ & $1$ & $\gtrsim 0.5$\cite{Moreno_2011, Ogata_1991} & $\sim \frac{2\pi v_F^2}{a|U|}$ & $\sim v_F$\\
     \hline
     $\frac{U}{\pi v_F} \ll -1, k_F=\frac{\pi}{2}$ & $1$ & none & $\frac{2\pi v_F^2}{a|U|}$ & none\\ \hline
\end{tabular}
\end{center}

\subsection{Numerical phase diagrams}

In this section, we show the behavior of a few ground-state observables that we used to extract the ground-state phase diagrams presented in the main text.
These observables map out different localization properties of the charge and spin sectors.

To quantify the presence or absence of non-Hermitian skin effect (NHSE) in each sector, we employ exact diagonalization of the non-Hermitian Hatano-Nelson-Hubbard model up to $L=16$ sites. 
We obtain the ground-state many-body wave function by performing an iterative Arnoldi procedure on a sparse encoding of the Hamiltonian, implemented in the \texttt{EDITH} software~\cite{EDITH}. 
Once the ground state is obtained, we calculate its density and spin profiles.

To measure localization of the charge sector, we compute a normalized mean center of mass (mcom) of the density profile,
\begin{equation}
\mathrm{mcom}_{\mathrm{charge}}
=
\frac{\displaystyle \sum_{j=0}^{L-1} 
j\, \big|\langle n_j \rangle\big|}
{\displaystyle (L-1)\sum_{j=0}^{L-1} 
\big|\langle n_j \rangle\big|}.
\end{equation}
This normalization ensures that $\mathrm{mcom}_{\mathrm{charge}}\in[0,1]$, where values close to $0$ ($1$) indicate localization near the left (right) boundary.

For the spin sector, since $\langle S_j \rangle$ can take both positive and negative values, we shift the profile by a constant offset $1/2$ before computing the center of mass,
\begin{equation}
\mathrm{mcom}_{\mathrm{spin}}
=
\frac{\displaystyle \sum_{j=0}^{L-1} 
j\, \big|\langle S_j \rangle + \tfrac{1}{2}\big|}
{\displaystyle (L-1)\sum_{j=0}^{L-1} 
\big|\langle S_j \rangle + \tfrac{1}{2}\big|}.
\end{equation}
The shift prevents cancellations between positive and negative spin contributions and provides a robust scalar measure of spin localization.

\begin{figure}
    \centering
    \includegraphics[width=\columnwidth]{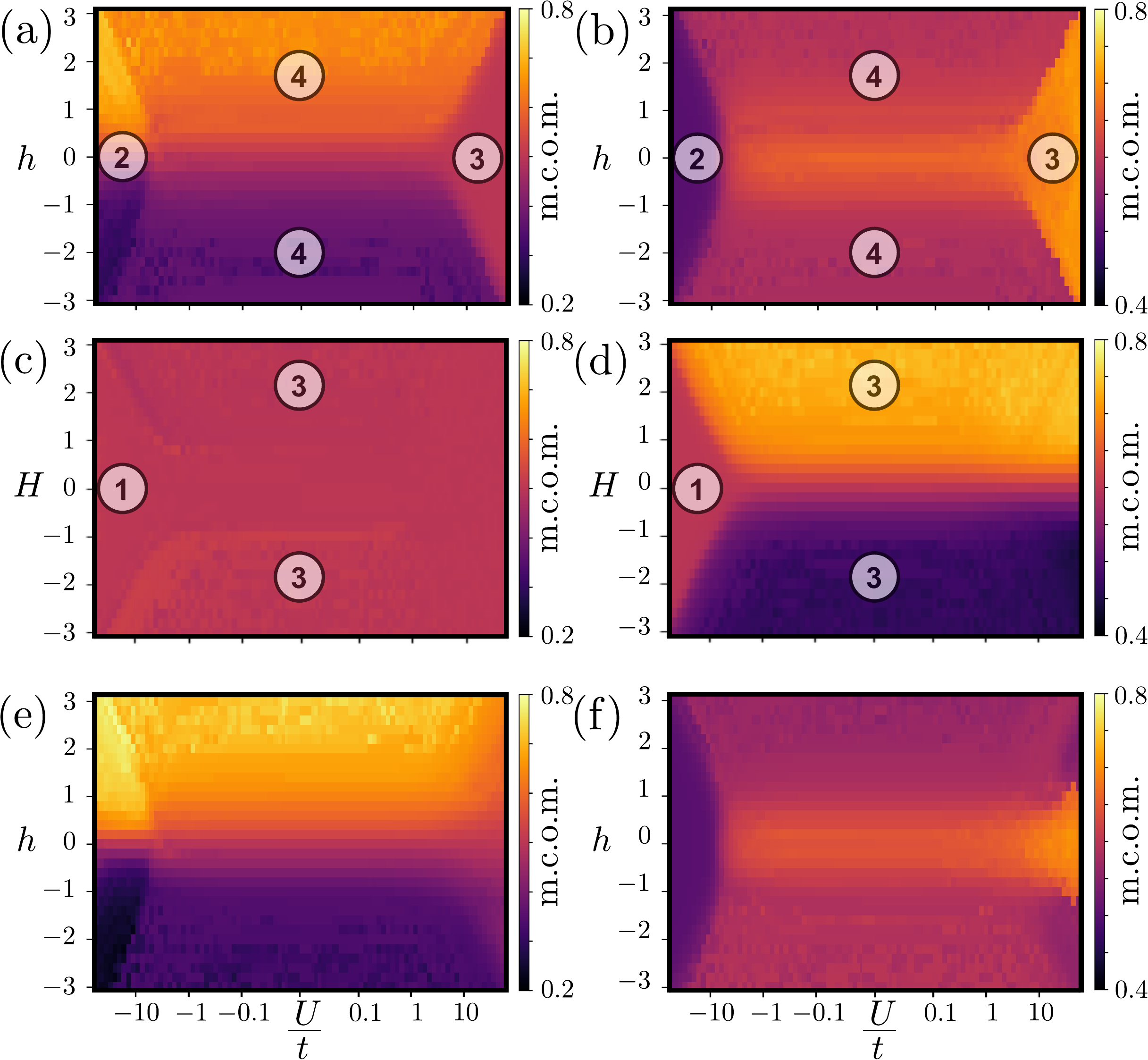}
    \caption{\textbf{Numerical phase diagrams for the Hatano-Nelson-Hubbard model}. Mean center of mass for the density (left panels) and the spin profile (right panels) in a non-Hermitian Hubbard model with $L=14$ sites, plotted as a function of interaction strength $U/t$ and gauge potential strengths $h$ and $H$.
    (a)-(b): mcom's for $H=1$ at half filling.
    (c)-(d): mcom's for $h=0$ at half filling.
    (e)-(f): mcom's for $H=1$ below half filling.
    The encircled numbers indicate regimes with qualitatively different localization behavior.
    }
    \label{fig:PD}
\end{figure}

The behavior of the charge and spin mcom for $L=14$ is shown in Fig.~\ref{fig:PD}. 
Panels~\ref{fig:PD}(a) and~\ref{fig:PD}(b) depict the mcoms at half filling as a function of $U/t$ and $h$ for a constant value of $H=1$. 
Panels~\ref{fig:PD}(c) and~\ref{fig:PD}(d) depict the mcoms at half filling as a function of $U/t$ and $H$ for a constant value of $h=0$. 
Panels~\ref{fig:PD}(e) and~\ref{fig:PD}(f) depict the mcoms below half filling (filling $\nu = 5/14$) as a function of $U/t$ and $h$ for a constant value of $H=1$.

The mcom plots reveal clear localization transitions in both charge and spin sectors as functions of interaction strength $U$ and gauge potentials $h, H$.
In regions where the mcom is close to $0.5$, the corresponding density remains approximately constant (delocalized) across the chain.
On the other hand, values approaching $0$ or $1$ signal boundary localization.
Moreover, we can clearly see that the sign of the gauge potentials controls the direction of localization in the sectors they couple to: positive (negative) values of $h$ localize the \emph{charge} sector to the right (left), while positive (negative) values of $H$ localize the \emph{spin} sector to the right (left).
Notably, there are extended regions where one sector is strongly localized while the other remains nearly uniform, demonstrating a clear separation of localization behavior between charge and spin.

At half filling, the combination of different localization features in the two sectors gives rise to four main regimes, labeled from (1) to (4) in Fig~\ref{fig:PD}.

Regime (1) does not display any type of NHSE and occurs when $h=0$ and for strong \emph{attractive} interactions $-U \gg t$.

Regime (2) is characterized by a skin effect in the charge sector only.
It also appears for strong attractive interactions, but for a nonzero value of both $h$ and $H$.

Regime (3) displays a skin effect in the spin sector only, and appears at strong \emph{repulsive} interactions $U \gg t$ when $H=0$, but can also emerge with attractive interactions if $h=0$.

Finally, regime (4) hosts a coexistence of skin effects in both sectors.
The accumulation of charge and spin can occur on the same side when the gauge potentials have the same sign, or at opposite ends when the gauge potentials have opposite sign.

Below half filling, the qualitative structure of the phase diagram remains similar, but two notable differences appear. 
First, the region corresponding to a pure spin NHSE becomes smaller, indicating that away from half filling the spin sector is more strongly influenced by charge fluctuations and less easily isolated as an independent localized mode. 
Second, localization features generally become sharper: the mcom values reach more extreme values closer to $0$ or $1$, signaling stronger boundary accumulation. 

\subsection{Double-sided skin effect below half-filling}
\begin{figure}
    \centering
    \includegraphics[width=0.7\columnwidth]{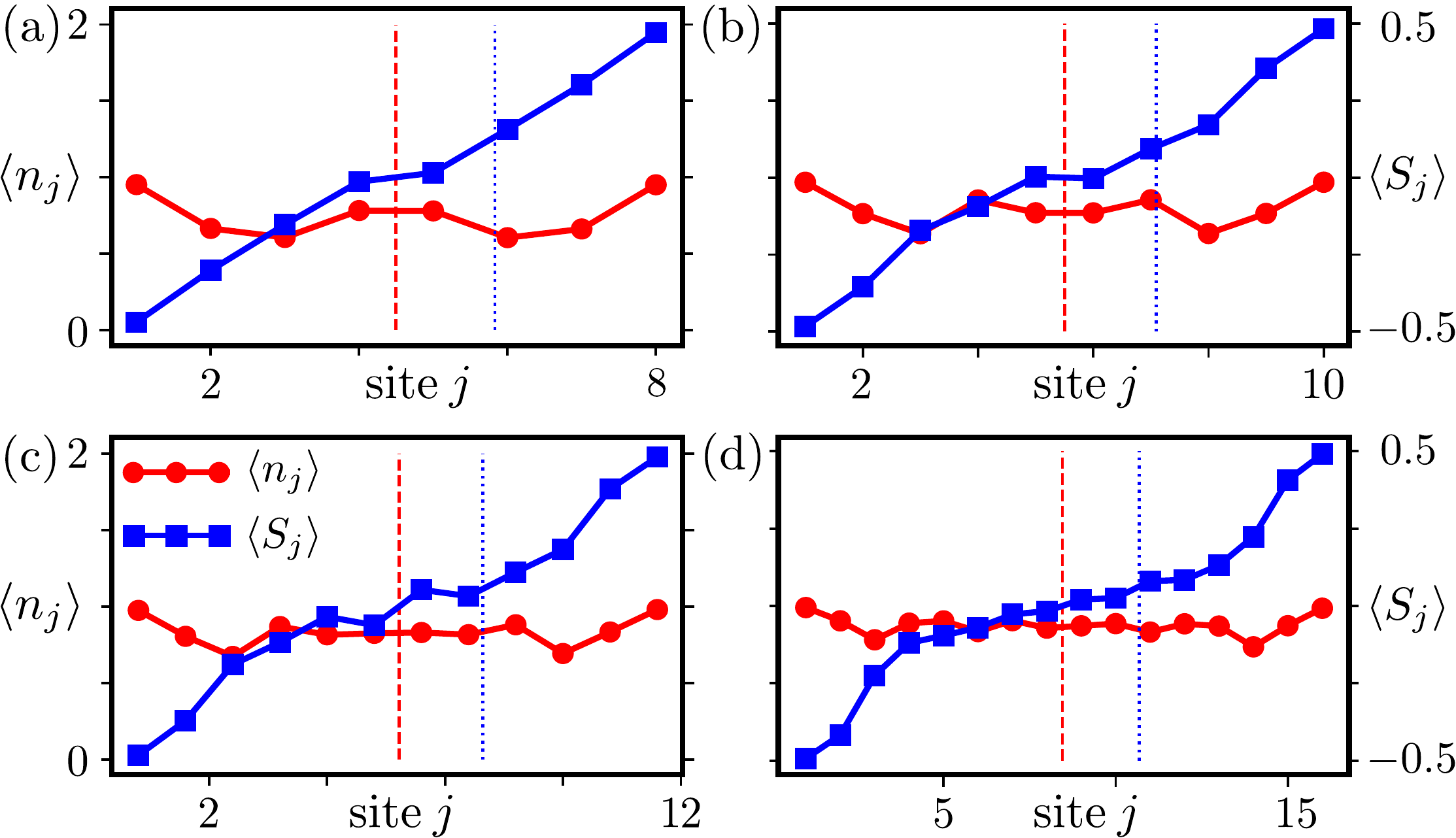}
    \caption{ \textbf{Numerical spin and charge profiles below half-filling.} Ground state spin and charge profiles below half-filling in the absence of an imaginary $U(1)$ gauge field. The systems sizes in the different panels are
    (a) $L=8$, (b) $L=10$, (c) $L=12$, (d) $L=16$.
    The vertical dashed lines indicate the mean centers of mass.
    All the other parameters are kept constant at $U/t=1$, $h=0$, $H=0.89588$.
    }
    \label{fig:double-edged}
\end{figure}

As described in the main text, below half filling irrelevant terms like $\lambda\partial_x\phi_c(v_{\sigma}^2(\partial_x\phi_{\sigma})^2-(\partial_t\phi_{\sigma})^2)$ are permitted by symmetry. Supposing that there is a static spin background, such as a spin skin effect, this may be expressed as $\sim \rho \sigma^2$, where $\rho=-\frac{1}{\pi}\partial_x\phi_c$ is the charge density and $\sigma=-\frac{1}{\pi}\partial_x\phi_{\sigma}$ is the spin density. Since this is effectively a chemical potential for the charge degree of freedom, a non-trivial spin profile in turn causes a non-trivial charge profile. This can also be seen numerically in the Hatano-Nelson-Hubbard model. In Fig.~\ref{fig:double-edged} we show spin and charge profiles for the Hatano-Nelson-Hubbard model below half filling for various system sizes in the presence of an imaginary $SU(2)$ vector potential, but without an imaginary $U(1)$ potential. As described in the main text, there is a spin skin effect (shown in blue), but the charges tend to accumulate slightly at the boundaries, too, due to the coupling between spin and charge below half filling. Furthermore, as the system size increases the double-sided charge skin effect becomes less pronounced, suggesting that it vanishes in the thermodynamic limit.

\subsection{Critical transition line}

We expect that the transition from a ground state with the non-Hermitian skin effect for some degree of freedom to one without occurs when that degree of freedom develops a gap. When this occurs can be determined from the field-theory model.
The low-energy effective field theory corresponding to the 1D Hubbard model at half-filling has the Hamiltonian
\begin{align}
    \mathcal{H}_B=\frac{1}{2\pi}\int dx&\ u_c\left(\frac{1}{K_c}\Pi_c^2+K_c(\partial_x\phi_c)^2+U_c\cos(2\sqrt{2}\phi_c)\right)+\nonumber\\
    +&u_{\sigma}\left(\frac{1}{K_{\sigma}}\Pi_{\sigma}^2+K_{\sigma}(\partial_x\phi_{\sigma})^2+U_{\sigma}\cos(2\sqrt{2}\phi_{\sigma})\right)
\end{align}
where, for weak interactions, $K_c\approx1+\frac{U}{\pi v_F}$ and $K_{\sigma}\approx 1-\frac{U}{\pi v_F}$. 
The scaling dimensions of the $U_i$ term are given by $\Delta_i=2K_i,$ and they are relevant whenever $\Delta_i<2$. 
This implies that the $U_c$ term is relevant when $U>0$ while $U_{\sigma}$ is relevant when $U<0$, so that a charge gap is opened for repulsive interactions and a spin gap is opened for attractive interactions. 
On a line, the spectrum is independent of the non-reciprocity, so that the transition from e.g. a ground state with a charge NHSE to a ground state with no charge NHSE happens at $U=0,$ and similarly for the spin NHSE. 
In the field theory description, if only relevant or marginal terms are taken into account, the spin and charge NHSE therefore never coexist at half-filling. The precise transition lines depend on irrelevant terms, such as band curvature as described in the main text, and the discrepancy between the field-theory and the lattice model may therefore be ascribed to such contributions.

We now turn to the full lattice model. 
To quantify the transition line numerically, we locate, for each fixed value of $h$, the sharpest variation of the density center-of-mass observable as a function of $U/t$.
We consider the observable $Z(U,h)$ (here the density mcom, although a similar procedure could be done with the spin mcom) and restrict to the positive-$U$ region, since the transition of interest occurs on the right-hand side of the diagram. 
To identify the point where the slope changes most abruptly, we compute
\begin{equation}
\frac{\partial Z}{\partial \log U},
\end{equation}
using a numerical gradient along the $\log U$ grid. 
Working in $\log U$ enhances sensitivity to sharp crossovers occurring at large $U$.
The critical interaction $U_c(h)$ is then defined as the value of $U$ at which this quantity is maximal. 
This procedure is fully automated, but each cut was also inspected by hand to verify that the selected point coincides with the visually identifiable change of slope in $Z(U)$.
The result of this procedure is a discrete set of transition points $\{ (h_i, U_c(h_i)) \}$ forming the boundary line.

\begin{figure}
    \centering
    \includegraphics[width=0.7\columnwidth]{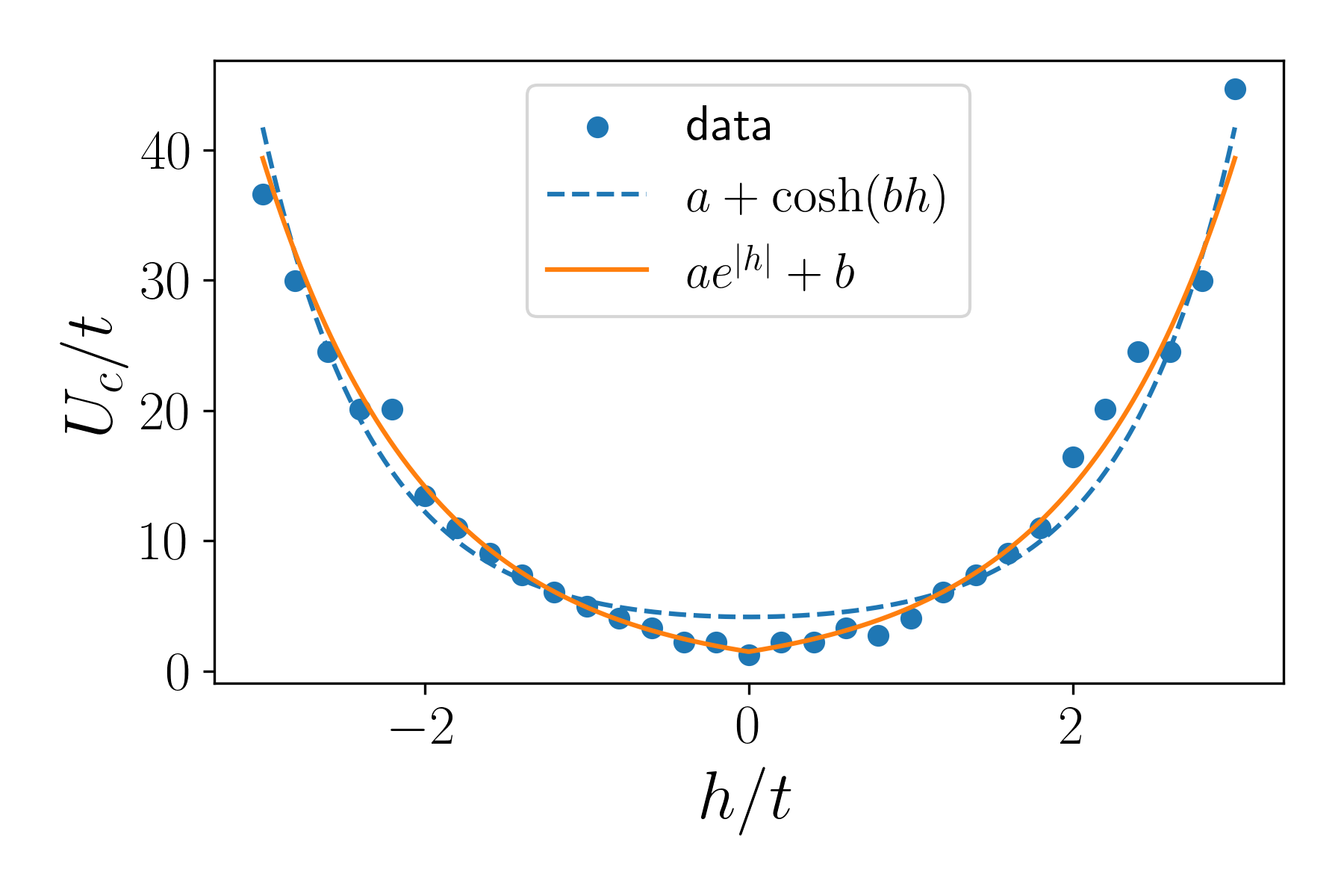}
    \caption{\textbf{Fit of the critical line separating pure spin NHSE from coexistence region.} Location of the critical interaction $U_c$ at which pure NHSE appears (diagonal line on the right-hand side of the phase diagram in the main text, Fig. 1), plotted as a function of gauge field $h$. The dots correspond to raw data extracted from exact diagonalization calculations. The curves correspond to two different kinds of fit -- an exponential fit (orange solid line)  and a cosine hyperbolic fit (blue dashed line).}
    \label{fig:fit}
\end{figure}

We tested two symmetric fitting ansätze for the full $h$ range:

\begin{align}
U_c(h) &= a + \cosh(bh), \\
U_c(h) &= a\, e^{|h|} + b.
\end{align}
Both capture the expected even dependence on $h$, as shown in Fig.~\ref{fig:fit}.
The fits were performed using nonlinear least squares over the full extracted dataset.

While the $\cosh$ form provides a reasonable interpolation at small $|h|$, it systematically underestimates the rapid growth of $U_c$ at large $|h|$. 
In contrast, the exponential form reproduces the large-$|h|$ behavior significantly better, yielding visibly smaller residuals and a more faithful description of the curvature of the boundary.

\subsection{Brief review of the $(E_8)_1$ current algebra}
In the main text, we construct a model exhibiting skin effects of charges corresponding to an $(E_8)_1$ current algebra. 
It consists of $8$ Cartan currents $H_j$ (corresponding to the generators of the maximal commuting subalgebra) and $240$ off-diagonal currents $V_{\alpha}$, each of the latter corresponding to a vector $\alpha$ in the root lattice of $E_8$ -- i.e. the lattice generated by all roots of the Lie algebra, which is 8-dimensional) \cite{cftbook}. 
Following \cite{Lim_2023}, the current algebra can be realized by $8$ free chiral bosons, $\Phi_j,\ j=1,2,...8$, where the Cartan generators are given by  $H_j=\partial_x\Phi_j$ and the off-diagonal currents by $V_{\alpha}=:e^{i\sum_j\alpha_j\Phi_j}:,$ which are vertex operators labeled by root vectors. The notation $::$ indicates normal ordering, and the root vectors $\alpha$ have integer entries and satisfy a norm condition given below. 

The Cartan matrix of $E_8$ encoding inner products between simple roots of the Lie algebra is given by
\begin{equation}
    K_{E_8}=\begin{bmatrix}
        2& -1 & & & & & & &\\
        -1&2&-1 & & & & & &\\
        & -1 & 2 & -1 & & & & &\\
        & & -1 & 2 & -1 & & & & \\
        & & & -1 & 2 & -1 & & -1&\\
        & & & & -1 & 2 & -1 & &\\
        & & & & & -1 & 2 & & \\
        & & & &-1 & & & 2
    \end{bmatrix}
\end{equation}
where only the non-zero elements were written out explicitly. 
Using this matrix, the two-point correlation functions of the chiral bosons corresponding to the $E_8$ theory given in the main text are 
\begin{equation}
    \langle\Phi_j(z)\Phi_k(w)\rangle =-(K_{E_8})_{jk}\log(z-w).
\end{equation}
where the complex coordinates $z=-i(x+vt),\ w=-i(x+vt)$ correspond to one chirality. The component of the field with the opposite chirality, which would be a function of $\Bar{z}=-i(-x+vt),$ has a similar two-point correlation function.
This fixes the normalization of the bosons and encodes their mutual couplings, which determines all remaining correlation functions in the theory. 
The root vectors $\mathbf{\alpha}$ satisfy the norm condition $\mathbf{\alpha}^TK_{E_8}\mathbf{\alpha}=2$ which ensures that the vertex operators $V_{\alpha}$ have conformal dimension $1$ (and hence correspond to conserved currents). 
The currents can then be shown to satisfy the algebra \cite{Lim_2023}
\begin{align}
    H_i(z)H_j(w)&\sim\frac{(K_{E_8})_{ij}}{(z-w)^2}\\
    H_i(z)V_{\mathbf{\alpha}}(w)&\sim\frac{(K_{E_8})_{ij}\alpha^j}{z-w}V_{\alpha}(w)\\
    V_{\alpha}(z)V_{-\alpha}(w)&\sim\frac{1}{(z-w)^2}+\frac{\alpha^i}{z-w}H_i(w)\\
    V_{\alpha}(z)V_{\beta}(w)&\sim  \frac{Z_{\alpha\beta}}{z-w}V_{\alpha+\beta}(w)
\end{align}
where $Z_{\alpha\beta}$ are a set of coefficients that ensure that $V_{\alpha}(z)V_{\beta}(w)=V_{\beta}(w)V_{\alpha}(z)$.

\subsection{More details on constructing the $(E_8)_1$ theory}
In the main text, the conformal embedding $U(11)_1=(E_8)_1\otimes U(3)_1$ is used to construct an interacting theory where the charges that accumulate at the boundaries due to the imaginary gauge fields are the Cartan charges of $(E_8)_1$. 
The decomposition is done explicitly in terms of chiral bosons in the following way (following and using the notation of \cite{Lopes_2019})
\begin{equation}
    \Phi_j^{\sigma}=\sum_{i\sigma'}U^{\sigma\sigma'}_{j i} \varphi_i^{\sigma'}
\end{equation}
where the matrix $U$ is integer valued to ensure that the vertex operators $V_{\alpha}=e^{i\sum_j\alpha_j \Phi_j}$ are local in terms of the physical fermions. It is given explicitly in \cite{Lim_2023}, but here only the fact that it is integer valued is of importance. The chiral bosons are related to the fields $\Phi,\phi$ in the main text as $\Phi_j=\frac{1}{2}(\Phi_j^++\Phi_j^-)$ for $j=1,2...,8$ and $\phi_{j-8}=\frac{1}{2}(\varphi^+_j+\varphi^-_j)$ for $j=9,10,11$, so that it splits the set of bosons into one sector generating $(E_8)_1$ and one generating $U(3)_1$.
\end{document}